\documentclass[a4paper]{mem}
\usepackage{natbib}
\usepackage{graphicx}
\usepackage[a4paper]{hyperref}
\begin{document}
\title{Twenty years of TIRGO telescope}

\author{F. Mannucci}
 %\inst{1}}

\offprints{F. Mannucci}
\mail{largo E. Fermi 5, I-50125, Firenze}

\institute{CNR, Istituto di Radioastronomia, sezione di Firenze,
largo E. Fermi 5, I-50125, Firenze}

\authorrunning{F. Mannucci}
\titlerunning{Tirgo telescope}

\abstract{
We present the characteristics and the current status of the
{\em Telescopio Infrarosso del Gornergrat (Tirgo)},
the national infrared facility operating from the beginning of the 80s
over the swiss alps. We describe its current instrumentation
and resume its activities during its 20 years of life.
}

\maketitle

\begin{figure*}
\centering
\resizebox{13cm}{!}{{\includegraphics{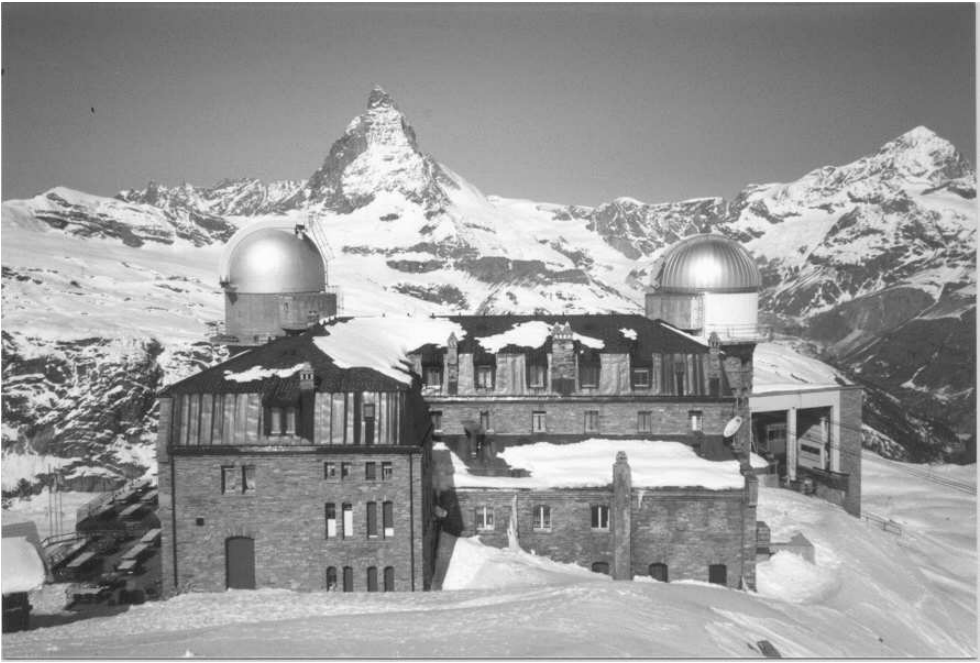}}}
\caption{\footnotesize{The domes of the Tirgo telescope (right)
and the submillimeter KOSMA telescope (left) as seen from the Gornergrat top.
On the background, the Mattelhorn and the Wieshorn. The sky is not always like
this.}}
\label{tirgo0}
\end{figure*}

\section{Introduction}

The Tirgo telescope has been operating for more than 20 years
at 3150 m of altitude on the Gornergrat mountain in the Swiss alps. It has a
1.5m primary mirror and is optimized for infrared astronomy, making it
one of the first 5 telescopes in the world to be capable of far-IR
observations.

\begin{figure}
\centering
\resizebox{\hsize}{!}{{\includegraphics{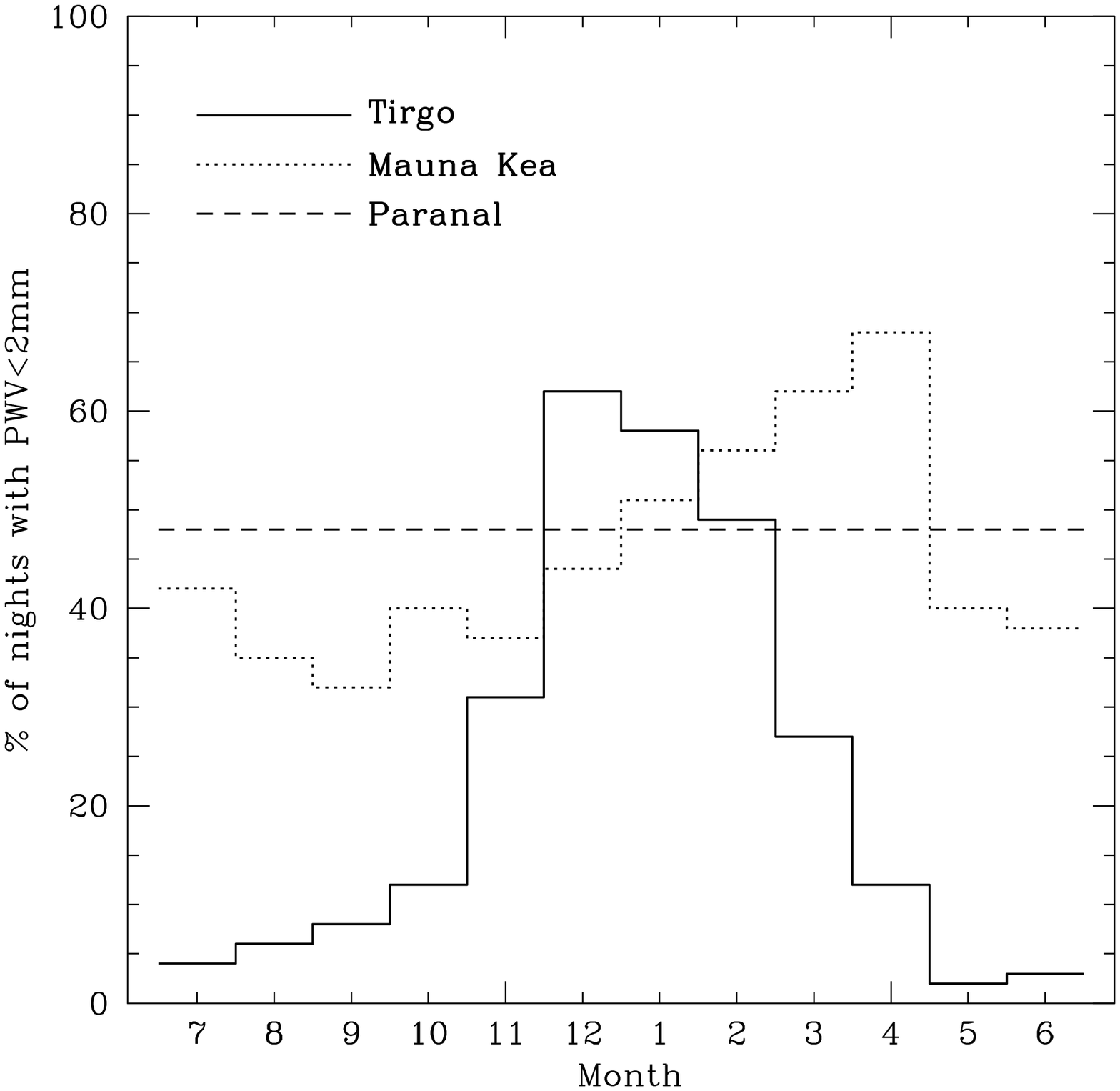}}}
\caption{\footnotesize{Average number of nights with precipitable
water vapor (PWV) below 2mm at Gornergrat, Mouna Kea and Paranal.
The value of 2mm was chosen as representative of a ``good'' night.
The Gornergrat data were taken from the KOSMA staff during 5 years
between January 1989 and September 1993 \citep{kramer}. Mouna Kea
and Paranal data are taken from the relative web pages. Only an
average over the year is available for Paranal. }} \label{pwv}
\end{figure}

\section{The site}

The top of the Gornergrat mountain (see Fig. \ref{tirgo0}) is one
of the highest location in Europe than can be reached in every
period of the year because of the presence of a rack-railway. It
was chosen as an astronomical site because during winter it has
low temperatures (between -10$^o$ and -20$^o$) and low
precipitable water vapor (PWV, see Fig. \ref{pwv}). During a few
tens of nights a year the conditions at Gornergrat are excellent
to allow for far-IR observations, and during this short time
Gornergrat is one of the best sites in the world. The drawback is
that the fraction of clear nights is usually around 35\% in winter
and even below during summer. For this reason the telescope in
usually closed during the 6 summer months and the total number of
observing nights is around 70 per year. The seeing in the
near-infrared bands is usually between 2 and 3 arcsec FHWM,
probably dominated by the local environment. These values of the
seeing and the low fraction of clear nights make Tirgo no longer
competitive with other national or international facilities for
near-infrared observations.

\section{The telescope}

Tirgo has a classical equatorial Cassegrain configuration, with a
1.5m primary. It is optimized for infrared observations, with no
buffles and a small (20 cm) secondary mirror that can oscillate up
to 30 Hz with a throw up to 5 arcmin. A ``cube'' mounted below the
primary mirror at the position of the secondary focus allows the
use of four instruments and an optical camera: a set of four
dicroics bend the infrared light to one of the four scientific
instruments while the optical light is collected by a camera for
pointing and tracking. Its is possible to switch from one
instrument to the other in just a few seconds.

Large efforts were spend in making the control of the telescope as friendly as
possible to reduce the assistance needed by the astronomers. Telescope and
dome set-up, pointing, tracking, focusing and offsetting are controlled by a
single program with an intuitive graphical user interface.
The same program can also handle target catalogs and is interfaced
with the acquisition software (specific to the detector in use) to pass
all the relevant informations as target name and telescope position.
By experience, any ``average'' astronomer can drive the telescope
by himself after half an hour of introduction.

\section{The instruments}

Many instruments were used at Tirgo (see Table \ref{table1}), most
of them developed specifically for that telescope. Several Italian
institutions were involved in this effort, in particular those in
Arcetri (Observatory, CNR and university), the CNR institutes of
IAS, IFSI, TESRE and IROE, and the Observatory of Turin. The
instruments currently available for public use are four:
\begin{itemize}
\item FIRT: near-IR single-element fast photometer. It is
sensitive between 1 and 5 micron using apertures between 7 and 40 arcsec.
Its fast electronics makes it possible a data rate up to 1 KHz. For this
reason FIRT is currently used mostly for lunar occultations (see below).
\item ARNICA: near-IR array camera. This is currently the most used
instrument. Its 256$\times$256 NICMOS3 detector is sensitive between 1 and 2.5
micron over a field-of-view of 4$\times$4 arcmin.
The camera is equipped with the standard JHKK$^\prime$ broad-band
filters, a set of narrow-band filters for line study and a set of
low-resolution grisms for spectra with R$\sim$300. The detection limits
(5$\sigma$ in 1 hour) are around 20.1 in J, 19.2 in H and 19.0 in K.
It was also mounted at several other telescopes: as WHT, SFOR, VATT, NOT and
TNG.
\item LONGSP: near-IR long slit spectrometer. It is based on the same array
as ARNICA and provides spectra with resolution between 1000 and 3000
along a 70 arcsec slit.
\item TIRCAM2: mid-IR array camera (see Persi, this volume).
Based on a Rockwell Si:As 128$\times$128 array, it is optimized for
observations around 10$\mu$m and contains several filters between 8 and
14$\mu$m. Its sensitivity (5$\sigma$, 1 hour) at 12.5$\mu$m is about 0.5 Jy.
\end{itemize}
The nitrogen needed for most of the infrared instruments is produces at the
telescope.

\section{Management, costs and publications}

Tirgo is an Italian national facility but is is open also to foreign
astronomers. A national time allocation committee is in charge of
reviewing the proposals twice a year and assigning observing time.

The institution in charge of the telescope is the
{\em Istituto di Radioastronomia, sezione di Firenze}
of the CNR, in collaboration with the {\em Osservatorio di Arcetri}
and with the {\em Universit\`a di Firenze}. {\em INAF} provides part of the
founding.
No one is working for the Tirgo telescope full-time, observations are
usually performed by a guest astronomer and one assistant from Arcetri,
rotating among a dozen astronomers and technicians.
The total amount of work needed for the observations and the maintenance
can be estimated in about 20 month/man per year, including
visiting astronomers. The cost of the operations are around 150 K-euro.

\begin{figure*}
\centering
\rotatebox[]{-90}{\includegraphics[width=7cm]{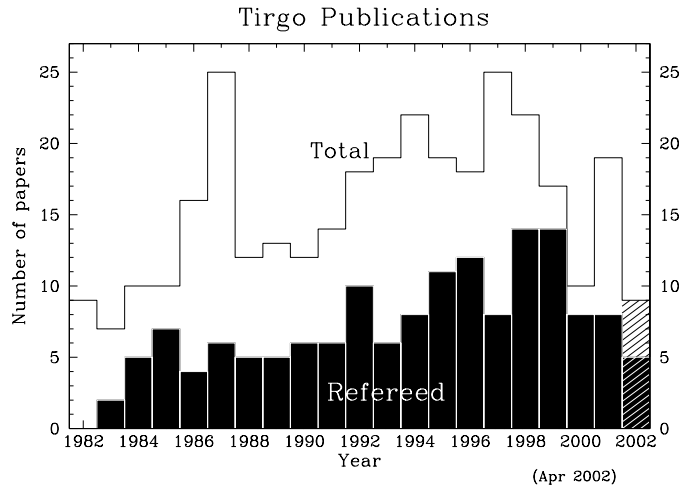}}
\caption{\footnotesize{Number of Tirgo publications from 1982. In black the
publications with a referee.}}
\label{papers}
\end{figure*}

Tirgo produced about 330 (known) papers, including 150 refereed
papers (see Fig. \ref{papers}). The rate of refereed publications
was about 5 per year until the introduction of ARNICA (1993), and
increased up to 14 afterward. In the last few years the rate is
decreasing because of the use of ARNICA at the TNG for two years
(1998 and 1999) and the availability of many near-IR cameras on
larger telescopes and in better sites.

\begin{table}
\caption{Instruments used at TIRGO (the list is probably
incomplete).}\label{table1}
\begin{tabular}{l}
\hline
near-IR InSb photometer           \\
mm GaGe photometer \\
Optical photometer         \\
mid-IR spectrometer \\
mid-IR camera TIRCAM \\
mid-IR camera CAMIRAS     \\
mid-IR bolometer \\
mid-IR camera TCMIRC      \\
far-IR bolometer \\
near-IR InSb photometer FIRT \\
near-IR spectrometer GOSPEC \\
near-IR camera ARNICA \\
near-IR spectrometer LONGSP \\
mid-IR camera TIRCAM2 \\
optical intensified camera  \\
optical CCD camera \\
800Ghz heterodine           \\
\hline
\end{tabular}
\end{table}

\section{Data Archive}

After one year of proprietary period, all the data taken from 1993 by
ARNICA and LONGSP becomes publicly available in the web site
html://tirgo.arcetri.astro.it/. A web form allows the selection
of the data from object name, target position, night of observation, filter
or file name. A total of about 330.000 images are available, 45GB of data.

\section{Some representative results}

Many scientific problems were addressed by Tirgo in 20 years of
observations. Here I list some of them to resume the scientific
activity at the telescope. The listed works are not necessary the most
important in their fields: the choice didn't follow any objective rule
but is mostly based of personal taste. This is also a sub-sample of the
results presented at the conference

\subsection{Lunar Occultations}

\begin{figure*}
\centering
\resizebox{12cm}{!}{{\includegraphics{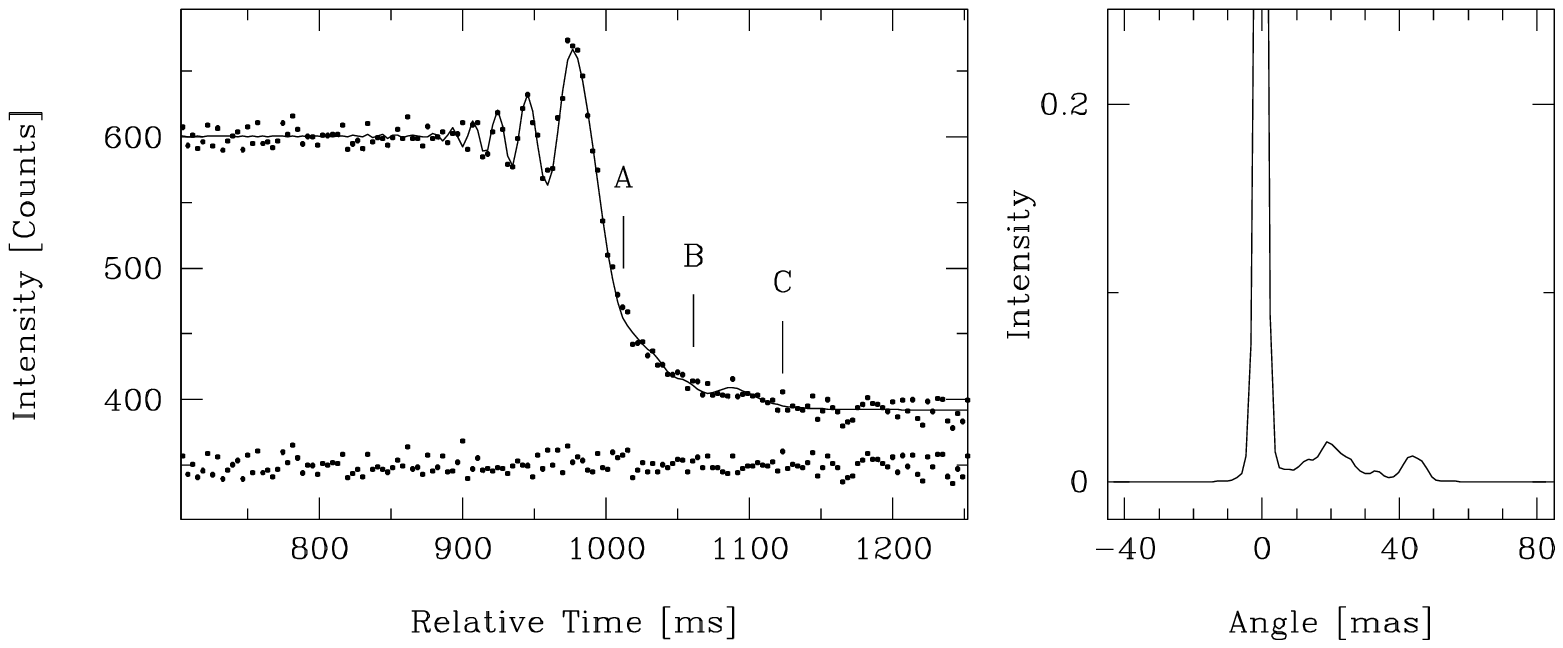}}}
\resizebox{12cm}{!}{{\includegraphics{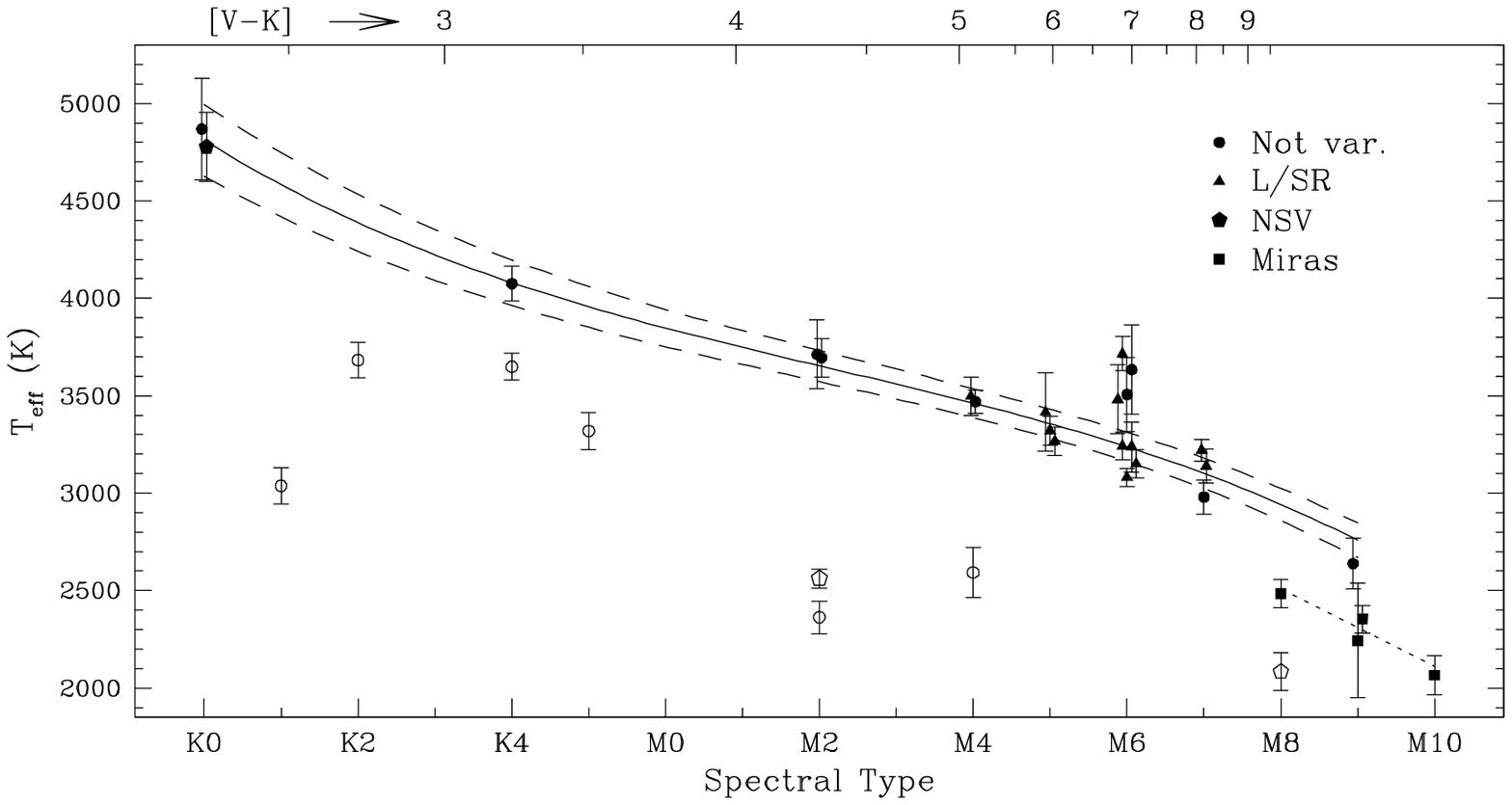}}}
\caption{\footnotesize{ {\em Upper panel}: lunar occultation of
SAO77810. The dots are the observation, the solid line a model
fit. The residuals are also shown. The source turns out to be a
triple star with the intensities shown in the right panel
\citep{richichi00}. {\em Lower panel}: Temperature scale of the
cold stars between classes K0 and M10 \citep{richichi99} as
measured using mostly radii and photometry obtained at Tirgo. The
solid line is the obtained mean calibration, the dashed lines
represent the range of associated error. }} \label{occult}
\end{figure*}

When a source is covered by the edge of the moon during its
motion, the diffraction pattern produced is a function of the
shape and the dimension of the source. Using sophisticated
deconvolution algorithms, stellar diameters as small as a few
milli-arcsec (mas) can be measured with precision of about 1 mas,
and the stellar multiplicity can be accurately tested. Lunar
occultations is one of the oldest Tirgo projects, started in
December 1985 and recorded more than 400 lunar occultations by
using both FIRT and ARNICA. The main scientific targets are the
measure of the frequency of binary stars, constraining models of
star formation, and the measure of the star diameter, a very
important parameter to study the stellar structure. Among the
results, the discovery of several tens of new binary and multiple
stars \citep[see e.g.][]{richichi02} and the measured of the
temperature scale of the cold stars \citet{richichi99}, (see Fig.
\ref{occult}).

\subsection{Comets}

\begin{figure*}
\centering
\resizebox{5cm}{!}{{\includegraphics{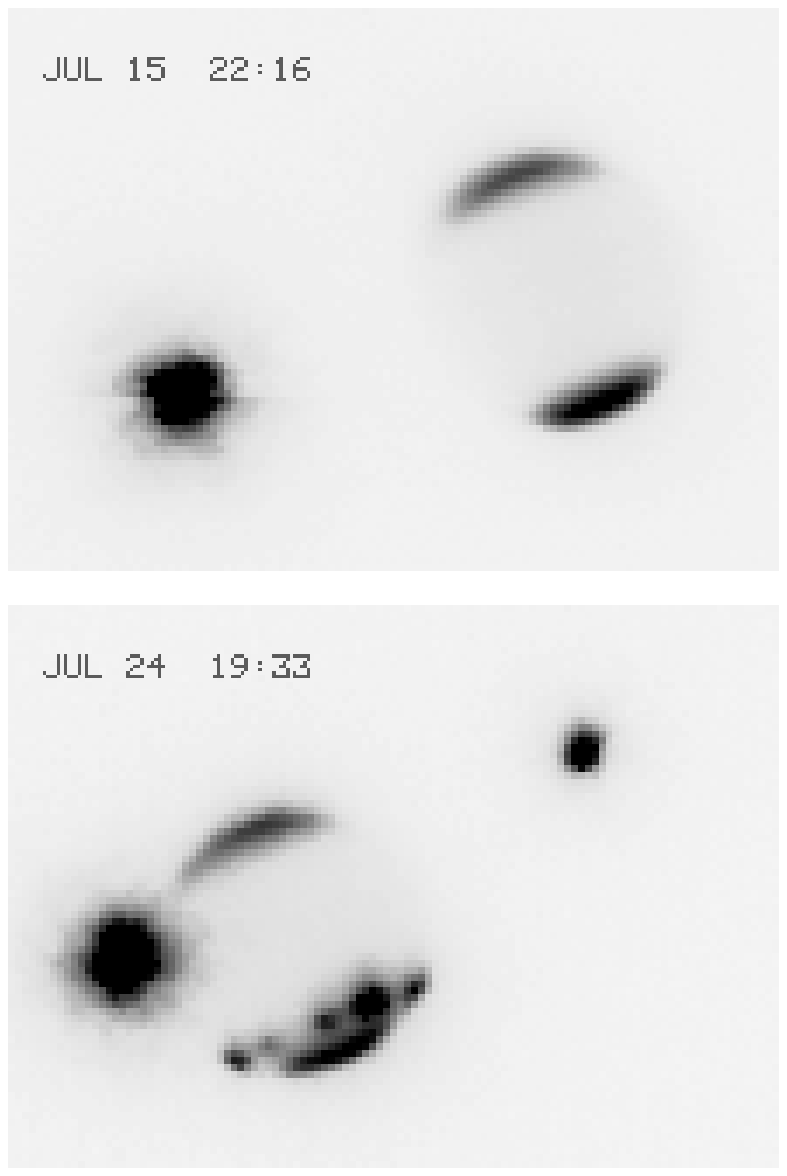}}}
\resizebox{8cm}{!}{{\includegraphics{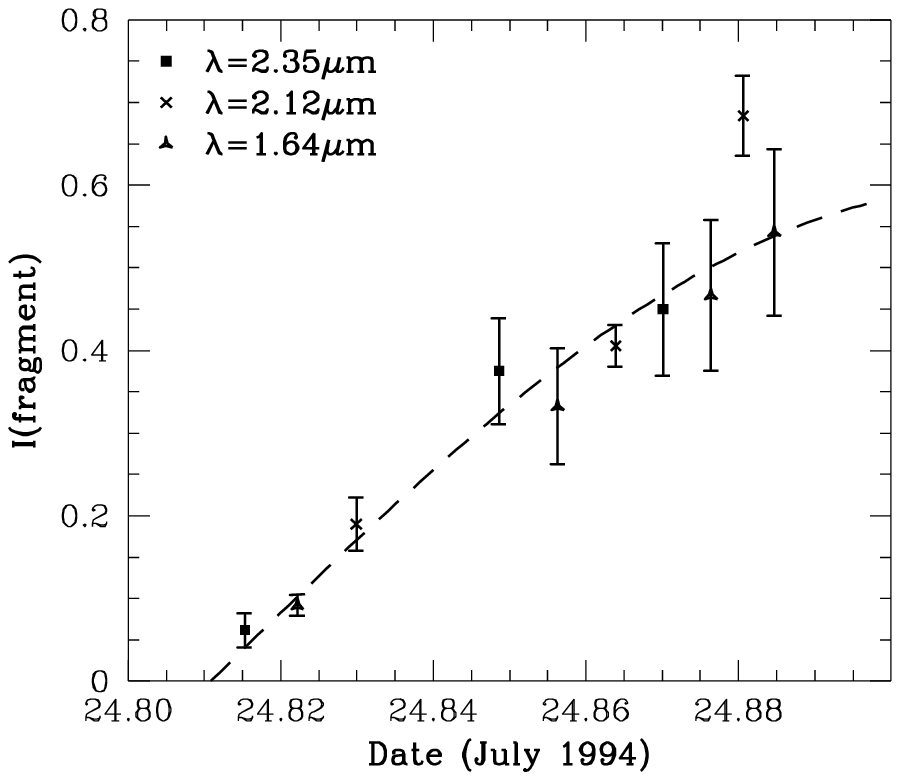}}}
\caption{\footnotesize{
{\em Left panel}: Jupiter before and after the collisions
with the fragments of the comet Shoemaker-Levy 9 in July 1994.
Before the impacts only the polar caps of the planet are visible because of
the use of a filter centered on a methane band. The bright spots outside
the planet circle are the satellite Io and Europa.
The loci of the impacts are visible near the southern
cap in the second image.
{\em Right panel}: the variation with rotation phase
of the brightness of the K+W fragment.
This evolution is well fitted by a simple
sin function with the expected values of phase and period,
indicating that the dust is geometrically thin and optically
thick. The albedo of the dust can also be measured and the results
support the hypothesis of the presence of silicate dust of 1$\mu$m grains.
}}
\label{jupiter}
\end{figure*}

Many comets, including SL9, Hyakutake and Hale-Bopp, were observed
at Tirgo with several instruments. The collision between Jupiter
and the comet Shoemaker-Levy 9 was observed in July 1994 using
ARNICA. A custom narrow-band filter centered on a methane
absorption band was used for this project. The atmosphere of the
planet is opaque at these wavelength and therefore in this filter
the planet appears dark, with some emission only from the polar
caps (see Fig. \ref{jupiter}). The fragments of the comet
deposited dust on the outer layers of the atmosphere and therefore
after the impacts these region appear bright due to the reflected
solar light. By using ARNICA observations \citet{tozzi} measured
the geometrical distribution of the dust and its albedo, giving
informations on the composition.

\subsection{Long wavelength observations}

In 1982 a GaGe bolometer was used at Tirgo to obtain observations
at 1mm of wavelength. \citet{mandolesi} observed the giant
molecular cloud W49 and detected it at the level of 1300 Jy. To my
knowledge this is the longest wavelength ever reached at Tirgo.

The second-longest wavelength published measures were obtained at
34 $mu$m between 1983 and 1988 by using a Ge bolometer
\citep{persi90}. The target was a sample of OH/IR stars observed
to derive the stellar mass loss rate and test the origin of the
pumping of the OH maser. The Tirgo observations between 2 and
34$\mu$m nicely fit the IRAS data.

\subsection{Imaging and spectroscopy of the Orion bar}

\begin{figure*}
\centering
\resizebox{6cm}{!}{{\includegraphics{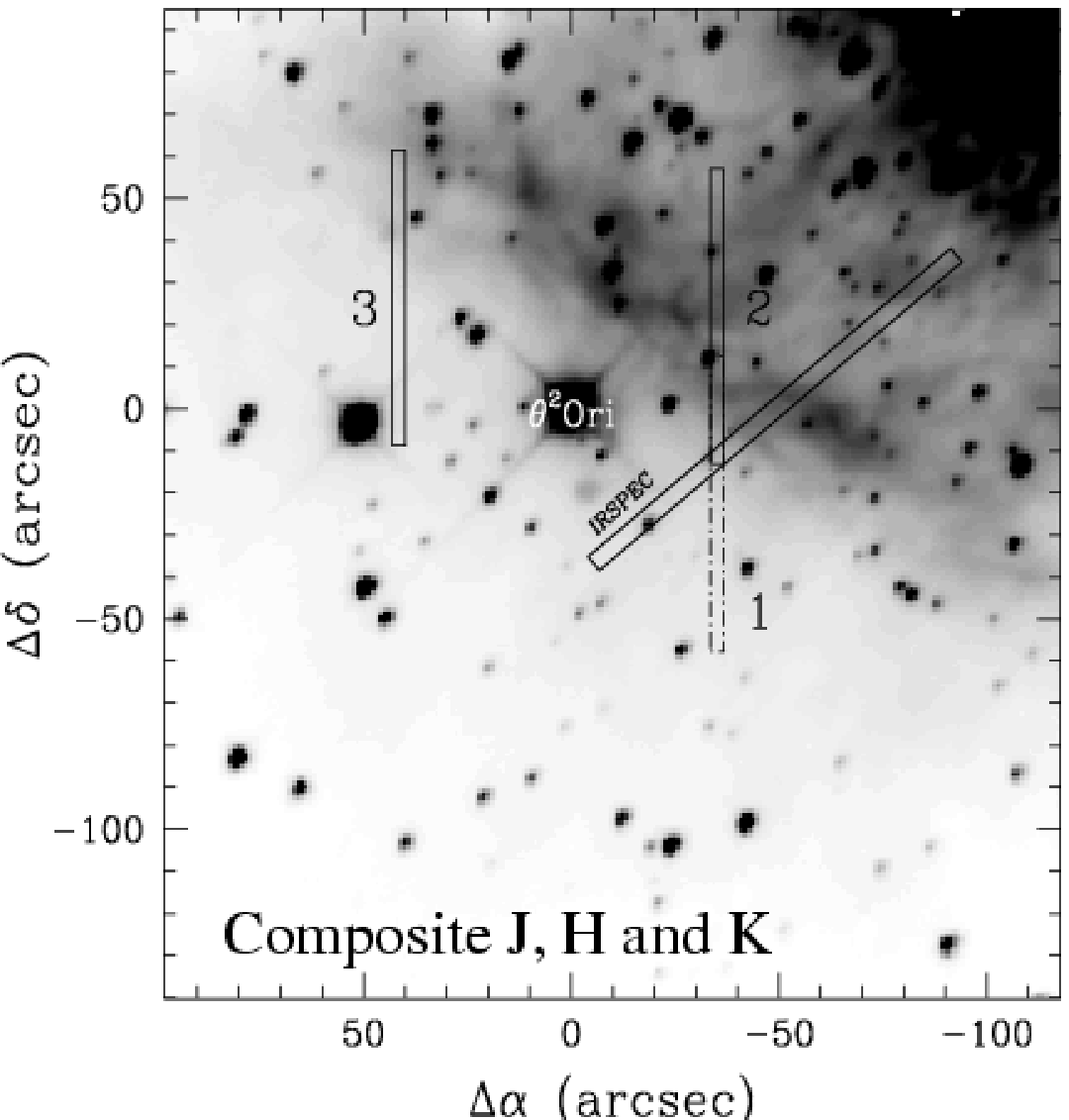}}}
\resizebox{7cm}{!}{{\includegraphics{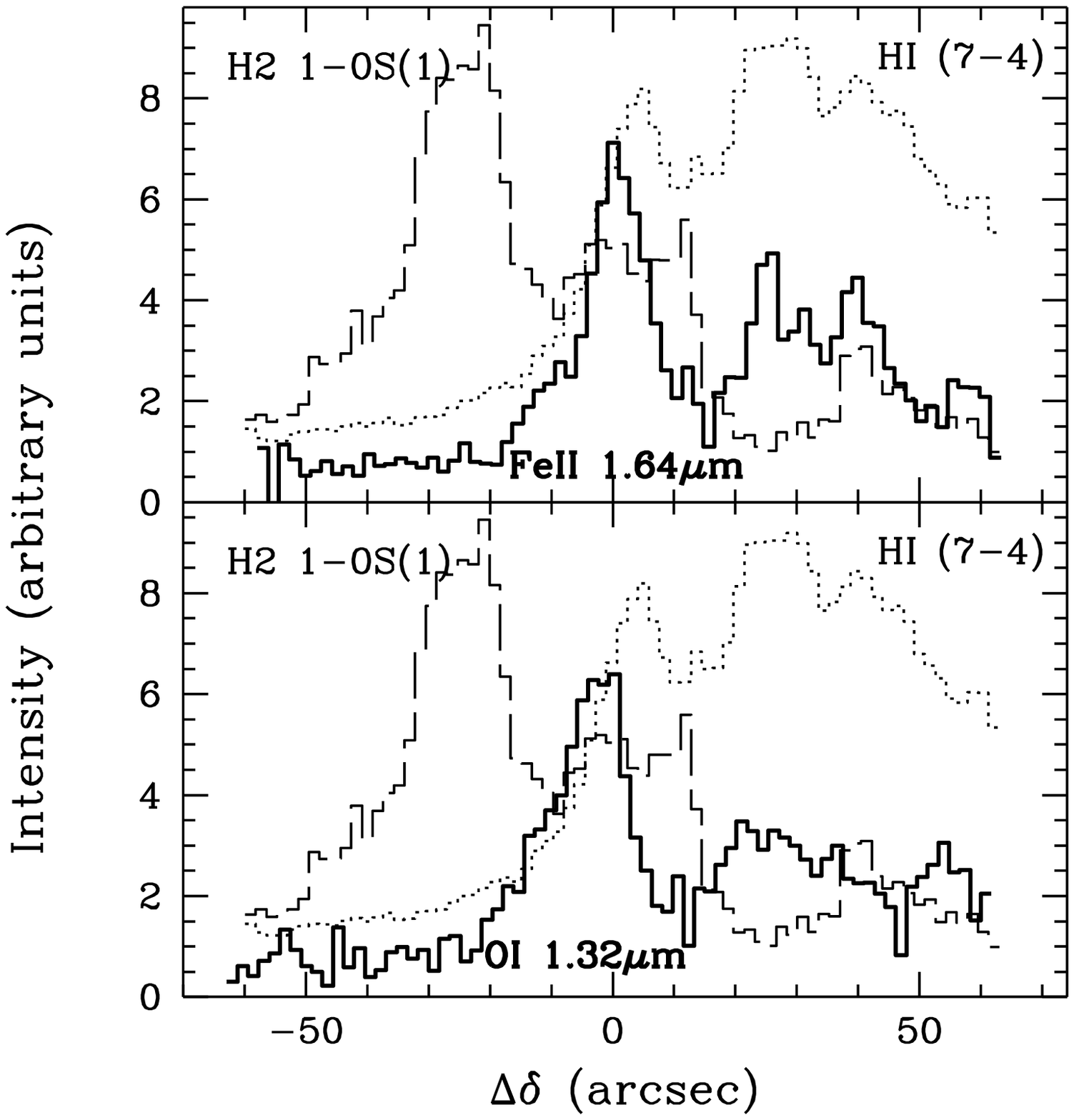}}}
\caption{\footnotesize{ARNICA and LONGSP observations of the Orion
bar from \citet{marconi}. {\em Left panel}: composite ARNICA J, H
and K image of the Orion bar. The positions of the LONGSP and
IRSPEC slits are indicated. {\em Right panel}: variation of the
brightness of various emission lines along the slit, tracing the
gas and radiation conditions across the region. Dotted line:
H(7-4); dashed line: H2 1-0S(1); solid line: FeII 1.64 $\mu$m in
the upper panel, OI 1.317 $\mu$m in the lower panel.. }}
\label{orion}
\end{figure*}

The Orion bar is one of the favorite target for infrared
astronomy, and Tirgo gave its contribution to the study of this
region of active star formation. \citet{marconi} used LONGSP to
observe this region and study the stratification of the emission
to derive density, temperature, geometric distribution and
radiation field in the various emitting regions (see Fig.
\ref{orion}) .

\subsection{Surface brightness of galaxies}

\begin{figure}
\centering
\resizebox{6cm}{!}{{\includegraphics{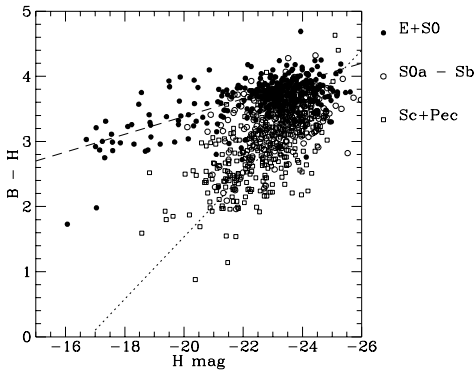}}}
\caption{\footnotesize{ Color magnitude relation for the galaxy in
the sample by \citet{gavazzi1,gavazzi2}. Black dots are
ellipticals and S0s, circles and squares are later type galaxies.
The two different behaviours are related to different formation
histories and star populations. }} \label{gavazzi}
\end{figure}

Gavazzi and co-workers have used ARNICA to observe over 900
galaxies of various morphological types in the H band.
Observations spanned three years from 1995 and 1997 and produced
the largest homogeneous sample on near-IR data of galaxies before
2MASS. The aim of this work was to measure the surface photometry
of a large number of galaxies to study several issues related to
the process of galaxy formation, as the color-magnitude relation
(see Fig. \ref{gavazzi}). As the mass-to-light ratio (M/L) in H
and K does not depend on galaxy luminosity, the near-IR bands are
in fact good tracer of the stellar mass.

ARNICA has quite a large field-of-view among the cameras based on
the 256$\times$256 arrays. This allows the observations of large,
nearby galaxies to study their detail properties. A large sample
of galaxies (about 200) were observed in J, H and K by Hunt,
Giovanardi, Moriondo and coworkers to deconvolve bulges and disks,
extract a nuclear point-like component, study the color gradients
due to both the stellar populations and to extinction effects,
study the global scaling relations for disks and bulges,
investigate the properties of the bars
\citep{moriondo98,moriondo99}.

\subsection{Spectra of normal galaxies}

LONGSP was used to observe a sample of large, nearby galaxies of
morphological type between E and Sc to define the first set of
template spectra on normal galaxies at near-IR wavelengths
\citep{mannucci}. 28 galaxies were observed in J, H and K using
apertures similar to those used by \citet{kinney} in the optical
to define their catalog of template spectra, allowing a reliable
matching of the two sets. The final uncertainties of the spectra
are between 1 and 3\%. These spectra are very useful to test the
galaxy spectrophotometric models which are usually calibrated by
using optical spectra only. The dominant stellar populations can
also be studied by the ratio between the equivalent widths of
several lines in the H and K bands.

\begin{figure*}
\centering
\resizebox{6.5cm}{!}{{\includegraphics{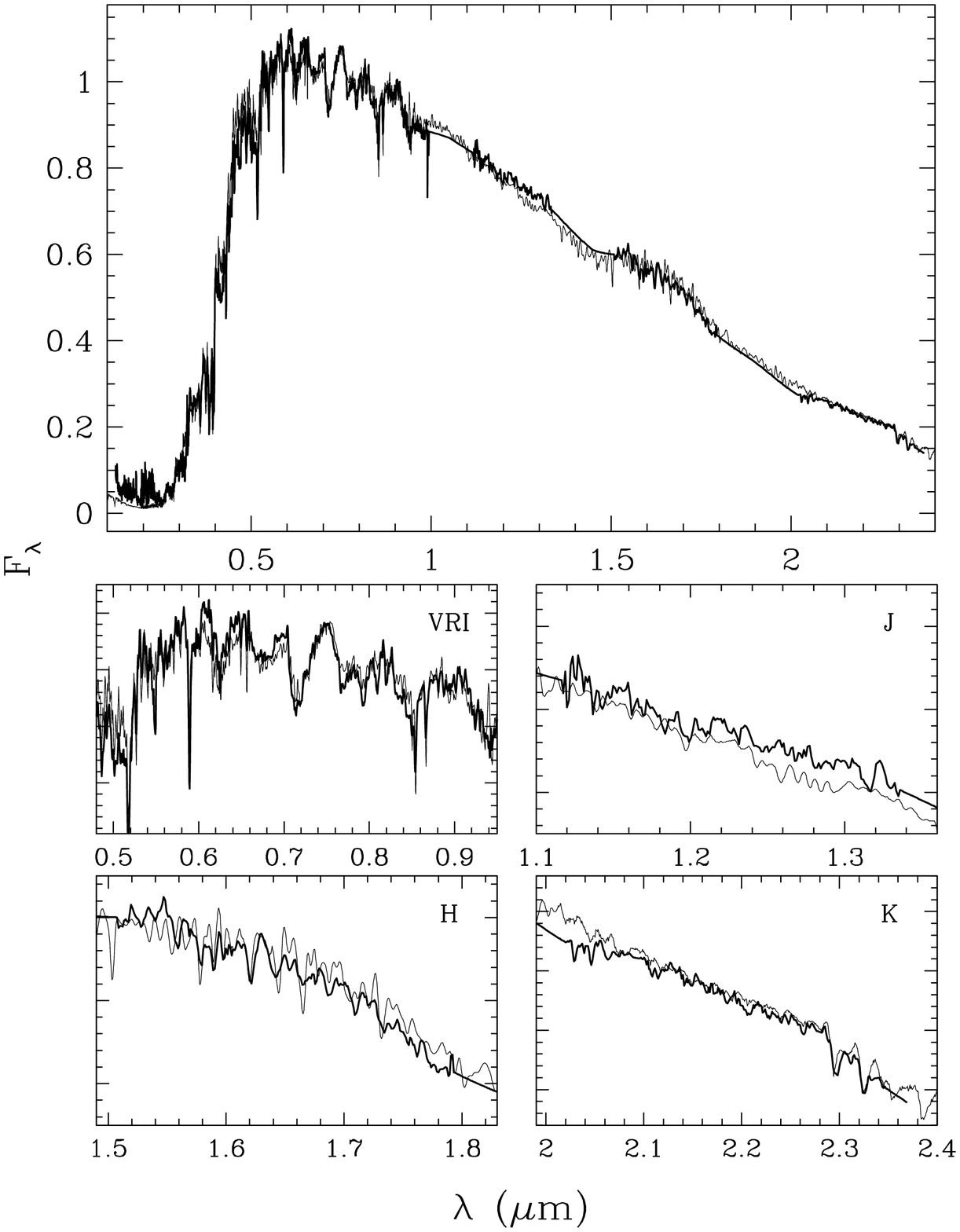}}}
\resizebox{6.5cm}{!}{{\includegraphics{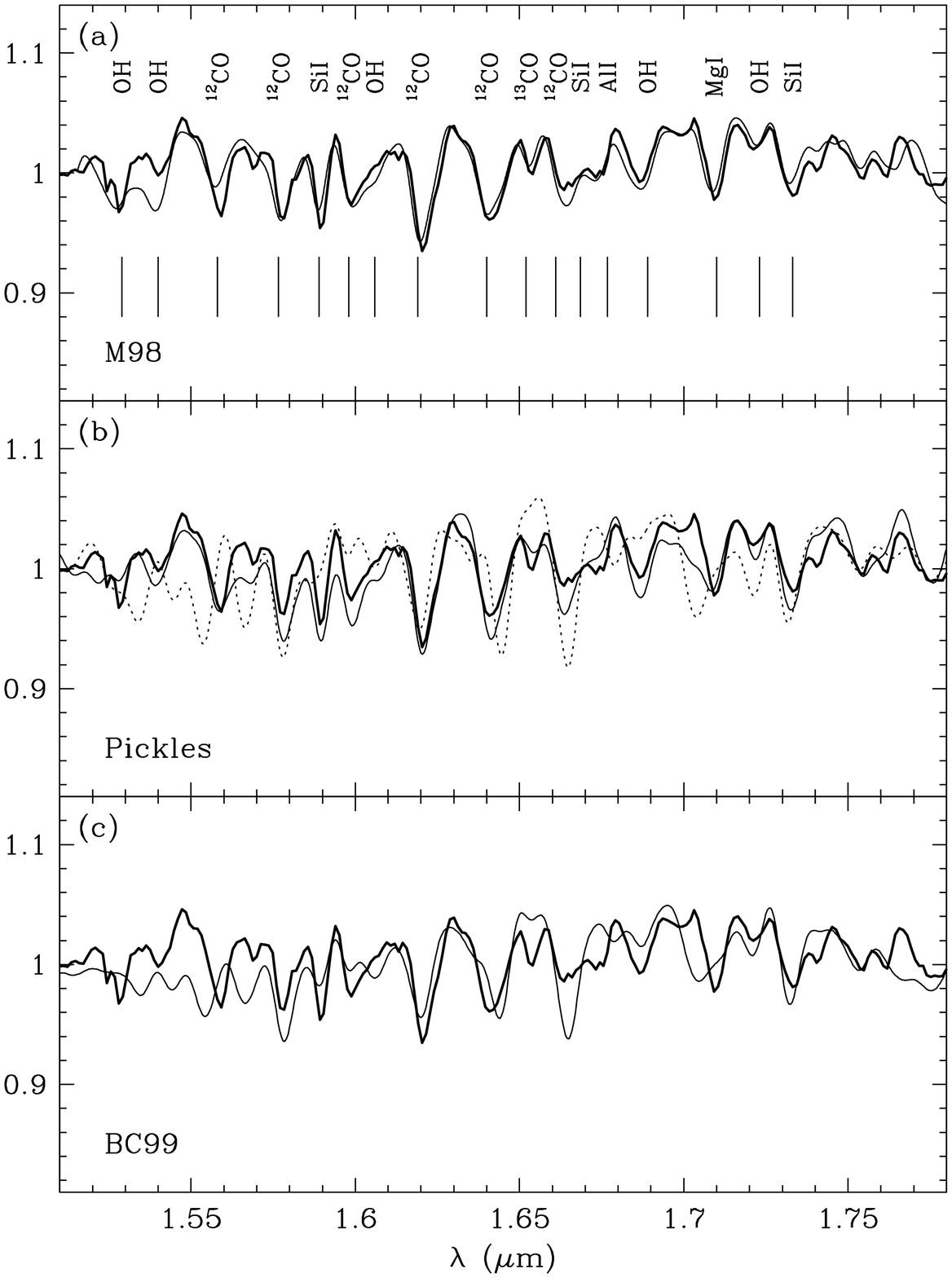}}}
\caption{\footnotesize{ {\em Left panel}: comparison between the
observed average spectrum of the elliptical galaxies (thick line)
with the prediction by \citet[][,BC99]{bruzual} model for a simple
stellar population 12 Gyr old. The overall spectral shape is very
well fitted, while many absorption lines are not correctly
reproduced. {\em Right panel}: Detail spectrum of the early-type
galaxies in the H band (thick line) compared with various
libraries of stellar spectra, \citet[][M98]{meyers},
\citet{pickles} and with the Bruzual \& Charlot spectrum in the
left panel. }} \label{mannucci}
\end{figure*}

\bibliographystyle{aa}

\end{document}